\documentclass[twoside]{article}

\usepackage{lipsum} 

\usepackage[sc]{mathpazo} 
\usepackage[T1]{fontenc} 
\usepackage{microtype} 

\usepackage[hmarginratio=1:1,top=32mm,columnsep=20pt]{geometry} 
\usepackage{multicol} 
\usepackage[hang, small,labelfont=bf,up,textfont=it,up]{caption} 
\usepackage{booktabs} 
\usepackage{float} 
\usepackage{hyperref} 

\usepackage{lettrine} 
\usepackage{paralist} 

\usepackage{abstract} 

\usepackage{titlesec} 
\renewcommand\thesection{\Roman{section}} 
\renewcommand\thesubsection{\Roman{subsection}} 
\titleformat{\section}[block]{\large\scshape\centering}{\thesection.}{1em}{} 
\titleformat{\subsection}[block]{\large}{\thesubsection.}{1em}{} 

\usepackage{fancyhdr} 
\usepackage[final]{graphicx} 

\usepackage[thinspace, thinqspace]{SIunits}
\usepackage[numbers]{natbib}

\title{\vspace{-10mm}\fontsize{24pt}{10pt}\selectfont\textbf{All-optical production and trapping of metastable noble gas atoms down to the single atom regime}} 

\author{M. Kohler$^1$, H. Daerr$^1$, P. Sahling$^1$, C. Sieveke$^1$, N. Jerschabek$^1$,\\ M.B. Kalinowski$^1$, C. Becker$^2$ and K. Sengstock$^2$\\
\large
\normalsize $^1$Carl Friedrich von Weizs\"acker Centre for Science and Peace Research,\\\normalsize University of Hamburg, 20144 Hamburg, Germany\\ \normalsize $^2$Institut f\"ur Laser-Physik, University of Hamburg, 22761 Hamburg, Germany\\ 
\vspace{-5mm}
}
\date{}
\begin{document}

\maketitle 

\thispagestyle{fancy} 


\begin{abstract}
{The determination of isotope ratios of noble gas atoms has many applications e.g. in physics, nuclear arms control, and earth sciences.
For several applications, the concentration of specific noble gas isotopes (e.g. Kr and Ar) is so low that single atom detection is highly desirable for a precise determination of the concentration.
As an important step in this direction, we demonstrate operation of a krypton Atom Trap Trace Analysis (ATTA) setup based on a magneto-optical trap (MOT) for metastable Kr atoms excited by all-optical means.
Compared to other state-of-the-art techniques for preparing metastable noble gas atoms, all-optical production is capable of overcoming limitations regarding minimal probe volume and avoiding cross-contamination of the samples.
In addition, it allows for a compact and reliable setup.
We identify optimal parameters of our experimental setup by employing the most abundant isotope $^{84}$Kr, and demonstrate single atom detection within a 3D MOT.
}
\end{abstract}


\begin{multicols}{2} 

\section{Introduction}\label{intro}
Tracers and their analysis are essential in earth sciences, physics, and nuclear arms control.
Natural and anthropogenic radioactive noble gas isotopes are ideally suited for use as tracers since they are chemically inert. 
Applications include Comprehensive Nuclear-Test-Ban Treaty verification based on xenon monitoring~\cite{kalinowski2006isotopic}, groundwater dating \cite{collon2004tracing, sturchio2004one}, and discovery of plutonium production (Kr) \cite{kalinowski2004conclusions}. 
However, the sensitivity of the aforementioned applications is limited by the analysis technique used.
Reduction of cross-contamination and sample size opens up many applications in the fields of research mentioned above.

One of the most profound and sensitive techniques is Atom Trap Trace Analysis (ATTA)~\cite{du2003realization}.
The basic idea is to count single atoms of the isotopes of interest within a mag\-ne\-to-optical trap (MOT), which enables measurements with a sensitivity down to 10$^{-12}$ for $^{81}$Kr~\cite{sturchio2004one}.

Since the optical transition for laser cooling and trapping from the ground state lies in the deep vacuum ultra violet (VUV) range, no suitable continuous wave laser source exists.
To overcome this problem, the noble gas atoms have to be prepared in a metastable triplet-state first (He, Ne, Ar, Kr, Xe,\cite{bardou1992magneto, shimizu1992double, katori1993lifetime, walhout1993magneto}).
For this purpose, state-of-the-art experiments use collisions between electrons and noble gas atoms in RF-driven plasmas, followed by a combination of Zeeman slower and MOT.

In these experiments the initial production of meta\-sta\-ble atoms is the limiting factor since a minimum gas density for sustaining the plasma discharge is required.
Further problems arise, for example, due to cross-con\-ta\-mi\-na\-tion of the sample of interest caused by krypton atoms implanted in the walls of the vaccum chamber during previous measurements \cite{lu2013tracer}. 
In subsequent measurements these atoms are released through ion and electron bombardment in the discharge plasma and contaminate the current sample. 
Therefore state-of-the-art ATTA set\-ups have to be cleaned for 36 hours using a xenon discharge to push the pre-implanted krypton out of the walls.
This limits the capacity per ATTA apparatus to about 120 samples per year \cite{lu2013tracer}.
Furthermore, analytical corrections depending on previous measurements have to be done to account for residual cross-contamination \cite{lu2010atom}. 
All-optical production of metastable noble gas has the potential to overcome these limitations \cite{lu2010atom}.

Here we report on the successful operation of a $^{84}$Kr and $^{83}$Kr 2D--3D MOT apparatus based on all-optical production of metastable atoms.
In addition, we demonstrate the single atom detection capability of our apparatus using $^{84}$Kr and $^{83}$Kr.

\begin{figure}[H]
 \resizebox{0.5\hsize}{!}{\includegraphics*{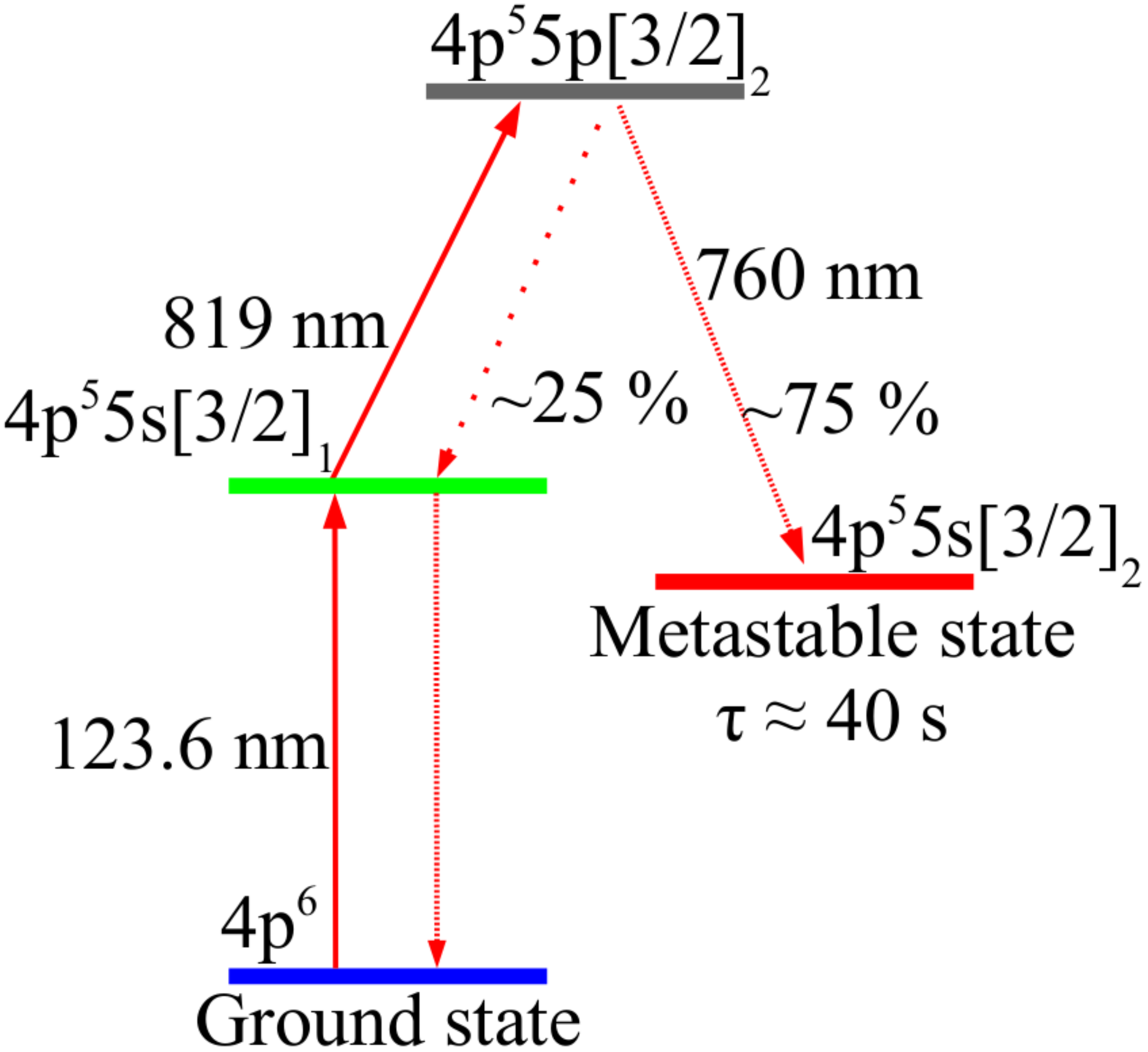}}
 \caption{Optical excitation scheme of krypton for transfer to the metastable state 4p$^5$5s[$3/2$]$_2$. \label{schemata}}
\end{figure}

\begin{figure}[H]
 \resizebox{0.5\hsize}{!}{\includegraphics*{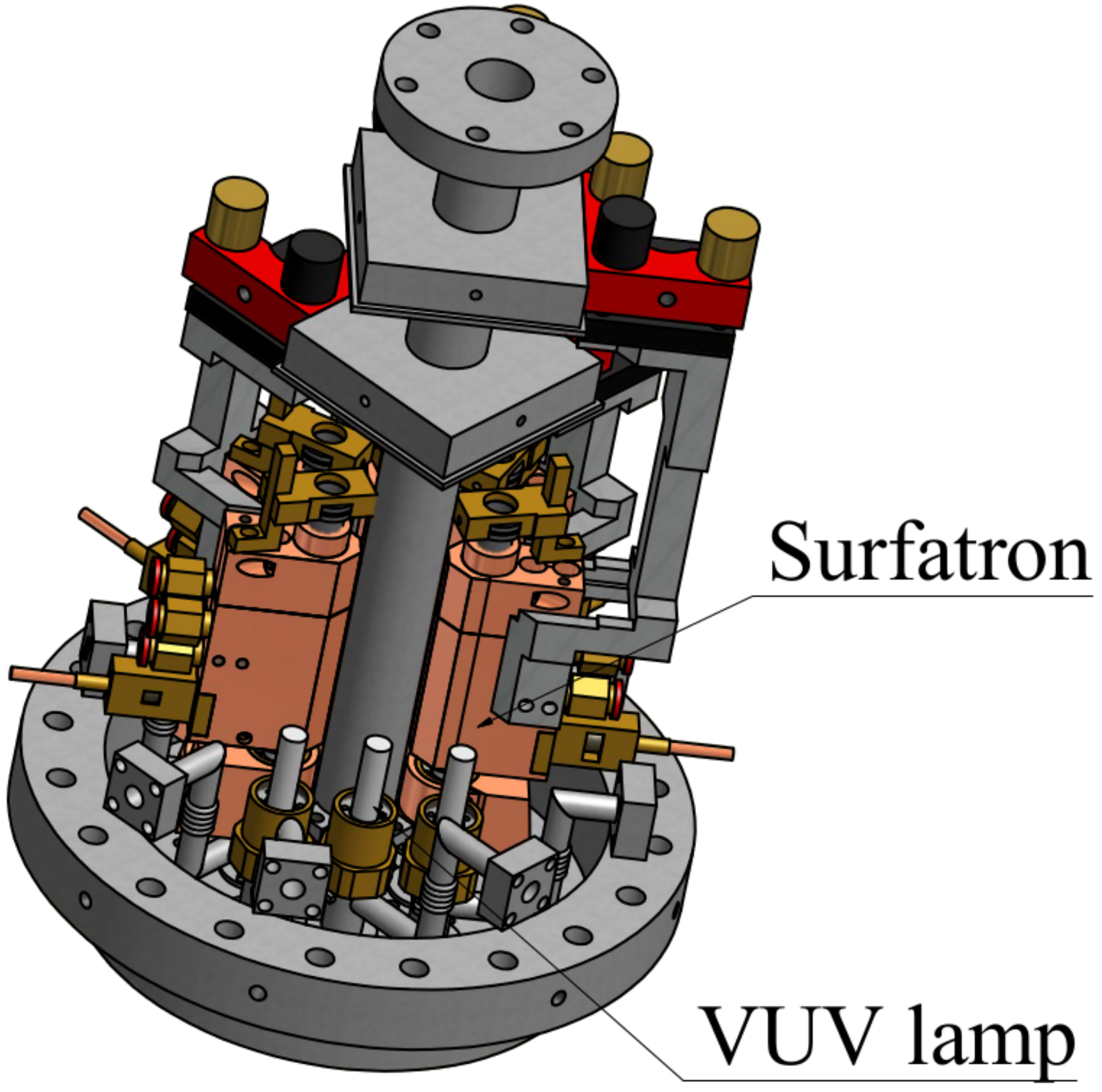}}
 \caption{Sketch of the VUV lamp setup attached to the top of the 2D MOT chamber.\label{deckel}}
\end{figure}

\begin{figure}[H]
 \resizebox{0.95\hsize}{!}{\includegraphics*{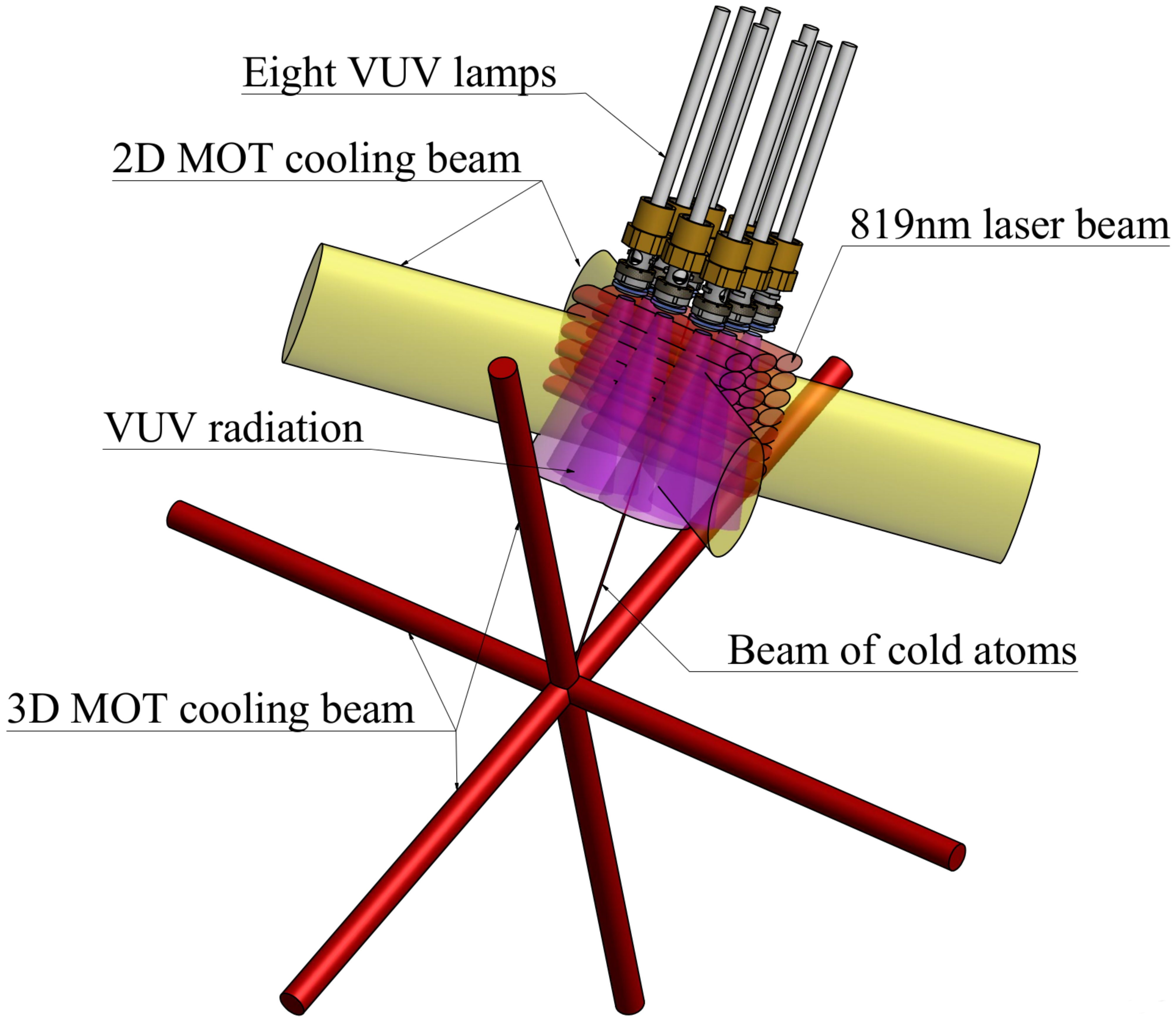}}
 \caption{
 Illustration of the beam guidance used in our ATTA setup.\label{aufbau} }
\end{figure}

The optical production of metastable krypton in the electronic configuration 4p$^5$5s[$3/2$]$_2$ (see Figure~\ref{schemata}) is based on the absorption of two resonant photons (\unit{123.6}{\nano\metre} \& \unit{819}{\nano\metre}), followed by spontaneous emission of a photon (\unit{760}{\nano\metre}) \cite{0953-4075-35-13-311}.
In our setup, the  radiation at \unit{123.6}{\nano\metre} is provided by an array of home built VUV lamps \cite{daerr2011novel} directly connected to the 2D MOT vacuum chamber (see Figure~\ref{deckel} \& \ref{aufbau}), while the infrared light is delivered by a \unit{1.4}{\watt} laser system. 
The final decay to the desired metastable state occurs with a probability of 75\%.


\section{Design and optical excitation}
Our new concept consists of a three chamber vacuum system;
a reservoir for storing and recapturing noble gas samples, and a 2D--3D MOT setup allowing for single atom detection.

The 2D MOT serves to achieve two important steps towards isotope selective single atom detection in our apparatus.
First metastable $^{84}$Kr is produced by all optical means using an array of self-made microwave driven VUV lamps \cite{daerr2011novel} and a transfer laser.
Subsequent laser cooling and trapping provides the required isotope selectivity.
The resulting beam of cold atoms is directed through a differential pumping stage (cone shaped type; smallest diameter \unit{1}{\milli\metre}) into a separate  3D MOT chamber, where the atoms are cooled and trapped.
We employ lock-in techniques (\unit{130}{\hertz}) to detect the fluorescence of numerous trapped atoms using a common photodiode; 
however for single atom detection an avalanche diode with continuous cooling of trapped atoms is implemented.

For experimentally achievable VUV intensity, the transition 4p$^6 \rightarrow$  4p$^5$5s[$3/2$]$_1$ is far from saturation. 
Hence, the number of metastable krypton atoms increases linearly with VUV intensity. 
Since the optical power of an individual lamp is limited and smaller than \unit{1}{\milli\watt} (\unit{5.5 \cdot\power{10}{14}}{}{Pho\-tons}/s \cite{okabe1964intense}) the only way to increase the VUV intensity is to use several lamps. 
We implemented an array of eight VUV lamps attached to the top of the 2D MOT chamber as depicted in Figure~\ref{deckel} \& \ref{aufbau}.

For an effective production of metastable krypton, the whole transfer volume under the lamps must be further illuminated with sufficient \unit{819}{\nano\metre} radiation.
This is done by a multiply saturated elliptic laser beam with diameters of \unit{10}{\milli\metre} and \unit{25}{\milli\metre} and a power of up to \unit{1.4}{\watt}, which is multi reflected by prisms under total reflection.

The 2D MOT, which is superimposed with the transfer volume, consists of a cylindrical quadrupole magnetic field and two retro reflected cooling beams to form a trapping volume that is as large as possible. 
The geometry of the 2D cooling beams has an elliptic shape with diameters of \unit{3}{\centi\metre} and \unit{7}{\centi\metre}.
The power of the beam on the symmetry axis corresponds to approximately 3 I$_\textrm{Sat}$. 
Four rectangular coils with 405 turns each are used to generate a cylindric quadrupole field with a magnetic gradient of up to \unit{1.45}{\milli\tesla\per\centi\metre}.

To minimize the loss of trapped Kr atoms through collisions with background atoms, the 3D MOT is realized in a separate vacuum chamber 
connected to the 2D MOT via a differential pumping stage allowing for a hundredfold lower pressure in the 3D MOT chamber.
To maximize the loading rate the distance between 2D and 3D MOT is kept as short as possible ($\sim$\unit{18}{\centi\metre}).

Six laser beams are used for 3D cooling and trapping, each having a diameter of \unit{30}{\milli\metre} and about \unit{20}{\milli\watt} of power.
The nearly spherical quadrupole magnetic field (\unit{0.8}{\milli\tesla\per\centi\metre} and \unit{1.75}{\milli\tesla\per\centi\metre}) is generated by two coils with 395 turns operated at \unit{2.5}{\ampere}.
\begin{figure}[H]
  \resizebox{1\hsize}{!}{\includegraphics*{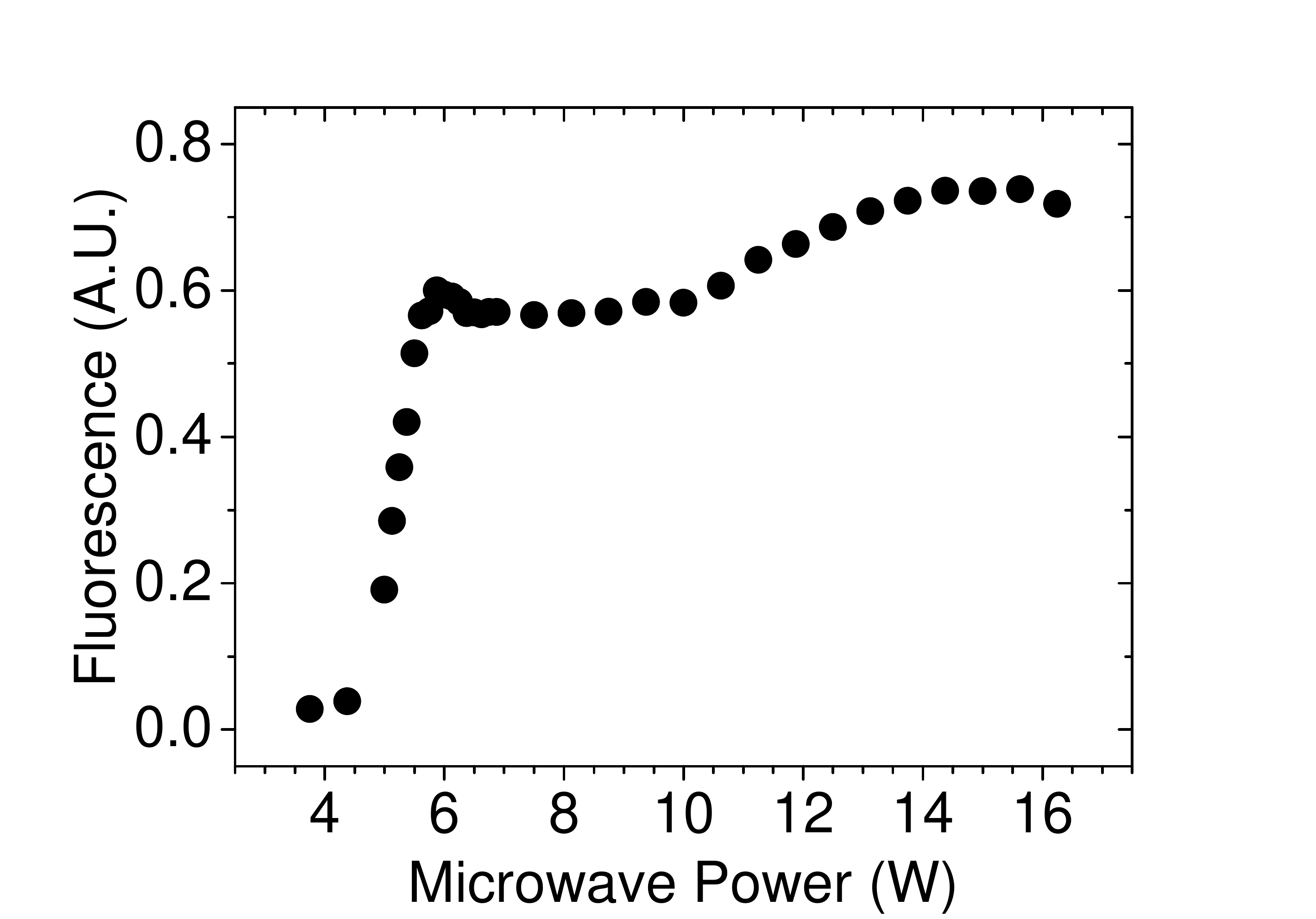}}
 \caption{Fluorescence of trapped atoms as a function of microwave power per lamp. 
 The experimental settings are $6 \cdot \unit{\power{10}{-6}}{\milli\bbar}$ 2D MOT pressure,  \unit{0.7}{\milli\tesla\per\centi\metre} 2D MOT magnetic field gradient, \unit{\sim 1}{\watt} transfer laser power and detunings of \unit{-2.5}{}$\gamma$ (2D), \unit{-1.5}{}$\gamma$ (3D).\label{microwave}
}
\end{figure}

\section{Characterization}

Various measurements have been carried out to characterize the experimental setup, both with $^{84}$Kr and $^{83}$Kr.
The following parameter optimization was performed with $^{84}$Kr.
We employed the fluorescence of the trapped atoms within the 3D MOT as a measure for the quality of the loading parameters of our apparatus.
In this way, we studied the effect of variations of microwave power, power of the \unit{819}{\nano\metre} radiation, pressure within the 2D MOT, and the current of the 2D MOT coils.

\textbf{Variation of microwave power}
One of the most crucial issues of all-optical production is the limited lifetime of the VUV lamps, mainly caused by color centers developing on the windows. 
This process mostly depends on the microwave power used to operate the lamps~\cite{daerr2011novel}.
Therefore, it is important to find the optimal point of operation in terms of balancing stability and microwave power.
For the experiments presented here, we typically use five lamps together with three ohmic loads.
We have measured the fluorescence of the 3D MOT as a function of microwave power per lamp between \unit{3.75}{\watt} and \unit{16.25}{\watt} and observed a steep increase between \unit{5}{\watt} and \unit{6.25}{\watt} (see Figure~\ref{microwave}).
This is the regime where the individual lamps start to sustain a stable plasma.
From this point on, the atomic fluorescence is constant up to approximately \unit{10}{\watt}.
Microwave power over \unit{10}{\watt} shows only a slight increase in fluorescence \cite{daerr2011novel}.

Under continuous operation we use a power of about \unit{11}{\watt} per lamp as a compromise between a sufficiently high fluorescence, stable operation, and an acceptable lifetime of the lamps.

\begin{figure}[H]
 \resizebox{1\hsize}{!}{\includegraphics*{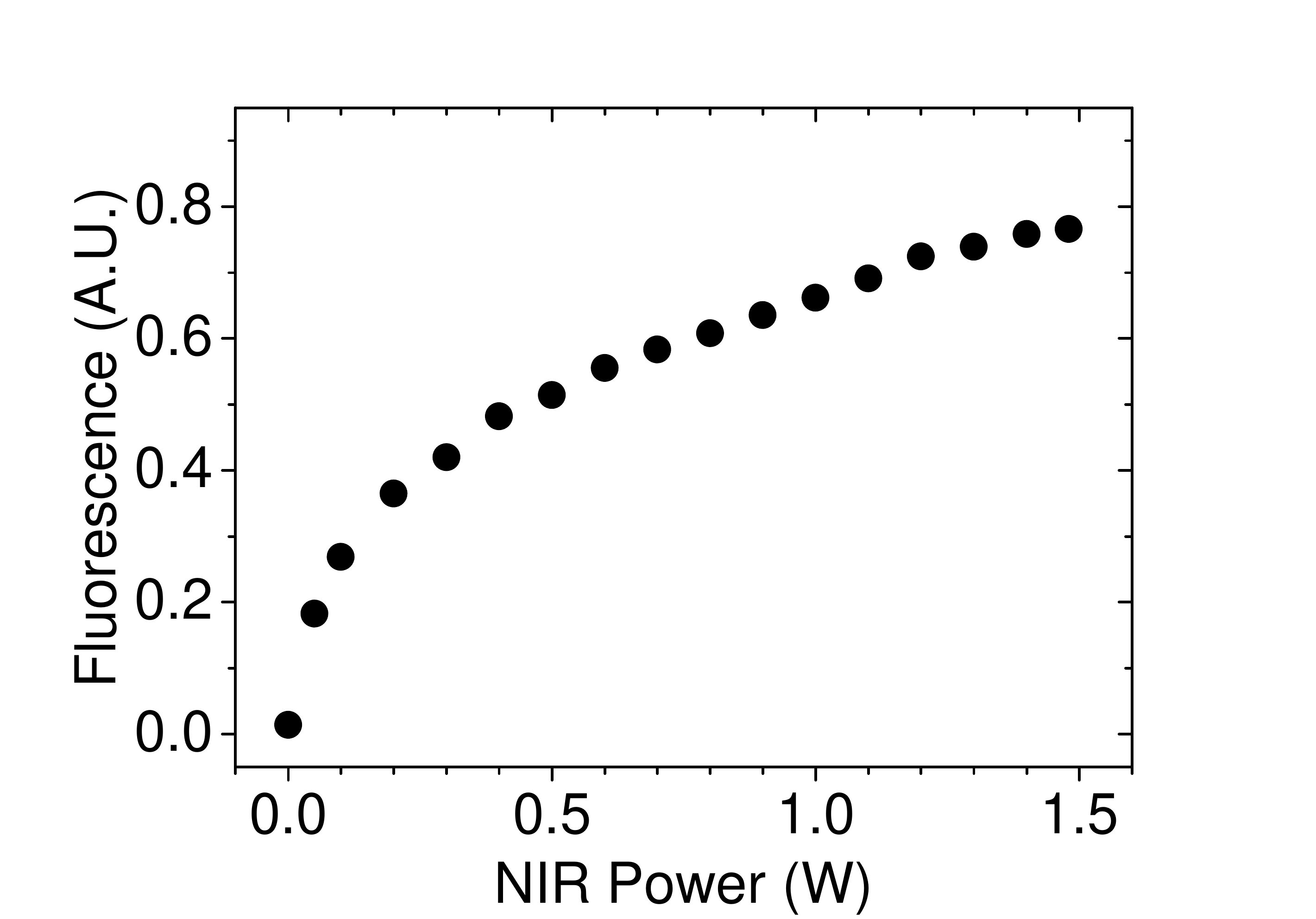}}
 \caption{Fluorescence of trapped atoms as a function of transfer laser power. 
 Experimental parameters are \unit{\sim11}{\watt} microwave power per lamp, \unit{0.7}{\milli\tesla\per\centi\metre} 2D MOT magnetic field gradient, $6.2 \cdot \unit{\power{10}{-6}}{\milli\bbar}$ 2D MOT pressure and detunings of \unit{-2.5}{}$\gamma$ (2D), \unit{-1.5}{}$\gamma$ (3D).\label{laser}
}
\end{figure}

\textbf{Power of the IR radiation}
In contrast to the power of the VUV lamps, there is enough power to saturate the transition 4p$^5$5s[$3/2$]$_1 \rightarrow\ $4p$^5$5p[$3/2$]$_2$ with \unit{819}{\nano\metre} radiation.
The efficiency of the all-optical excitation as a function of the \unit{819}{\nano\metre} radiation power is depicted in Figure~\ref{laser}.
Note that we used a combination of two separate laser sources for measurements exceeding \unit{1.0}{\watt}.
As a consequence, the fluorescence exhibits a discontinuity as a result of a marginal mismatch of these two laser beams.
In accordance with these measurements, we typically work with \unit{1.4}{\watt} at \unit{819}{\nano\metre}. 

\begin{figure}[H]
  \resizebox{.95\hsize}{!}{\includegraphics*{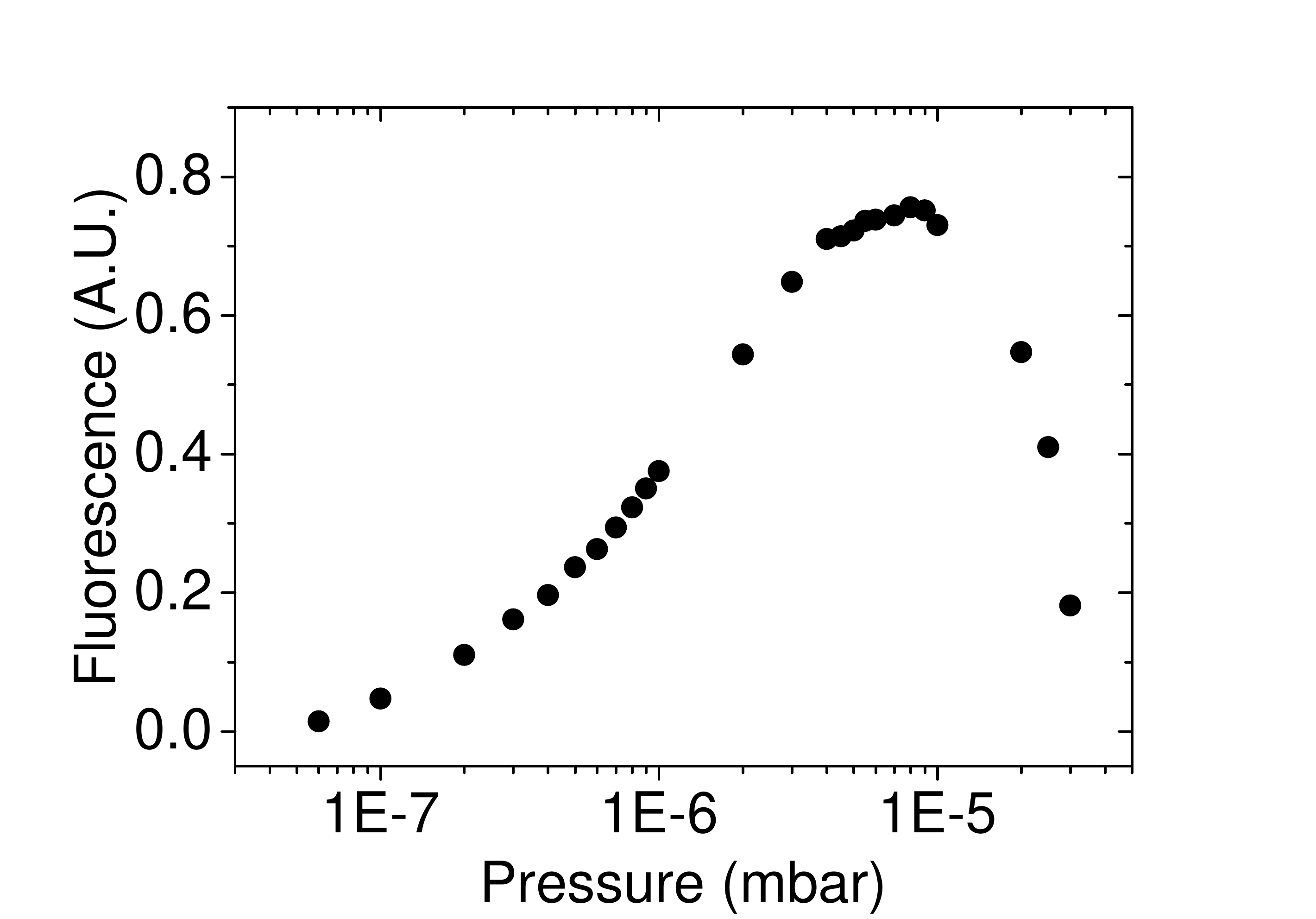}}
 \caption{Fluorescence of trapped atoms as a function of pressure within the 2D MOT chamber.
  Experimental parameters are \unit{\sim11}{\watt} microwave power per lamp, \unit{0.7}{\milli\tesla\per\centi\metre} 2D MOT magnetic field gradient, \unit{1.4}{\watt} transfer laser power and detunings of \unit{-2.5}{}$\gamma$ (2D), \unit{-1.5}{}$\gamma$ (3D).
\label{druck}
}
\end{figure}

\textbf{Krypton partial pressure of the 2D MOT}
Optimization of the Kr partial pressure in the 2D MOT allows us to achieve the highest flux of cold atoms being fed to the 3D MOT, enabling reasonably fast measurements.
Two contrary effects dominate the pressure dependence of the fluorescence in the 3D MOT.
First, the production rate of metastable Kr atoms, which is proportional to the Kr partial pressure, benefits from increased pressure.
Second, there are various loss mechanisms, such as collisions with background atoms, as well as collisions among meta-stable slow Kr atoms and ground state Kr atoms in the 2D MOT.
Furthermore, in the 3D MOT, background gas collisions, inelastic two-body collisions among metastable Kr atoms, as well as radiative decay of the metastable state and the non-unity duty cycle of the lock-in detection lead to additional loss.
In general, the different loss rates depend on the Kr partial pressure in the 2D MOT either linearly or quadratically, hence giving rise to an optimum in 3D MOT atom number as a function of the pressure.
Note that the background pressures of the 2D and 3D MOT are linearly coupled through the differential pumping stage.
Our measurements shown in Figure~\ref{druck} reveal a behavior that confirms these qualitative considerations. We observed the highest fluorescence at about \unit{8\cdot \power{10}{-6}}{\milli\bbar} .

\begin{figure}[H]
 \resizebox{1\hsize}{!}{ \includegraphics*{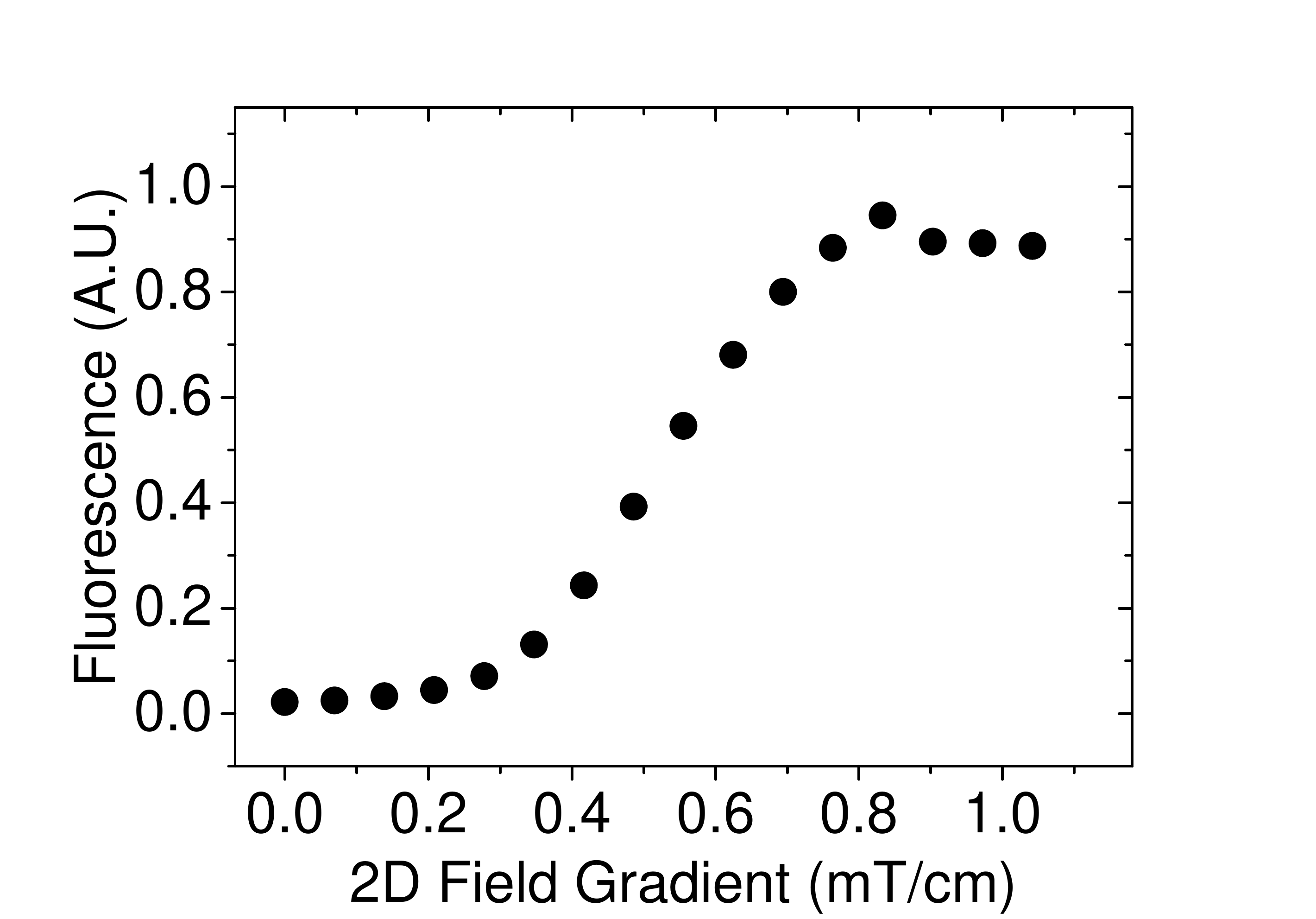}}
 \caption{Fluorescence of trapped atoms as a function of 2D MOT magnetic field gradient.
  Experimental parameters are \unit{\sim11}{\watt} microwave power per lamp,  $6.2 \cdot \unit{\power{10}{-6}}{\milli\bbar}$ 2D MOT pressure, \unit{1.4}{\watt} transfer laser power and detunings of \unit{-2.5}{}$\gamma$ (2D), \unit{-1.5}{}$\gamma$ (3D).\label{strom}
}
\end{figure}

\textbf{Magnetic field gradient of the 2D MOT coils}
The characterization of the 2D MOT includes determining the dependency on the magnetic field gradient.

At about \unit{0.7}{\milli\tesla\per\centi\metre} 2D MOT coil current, the positions of the coils were optimized to maximize the fluorescence signal of trapped atoms.
Subsequently, the dependency of the fluorescence on the magnetic field gradient of the 2D MOT was determined.
In Figure~\ref{strom} the fluorescence of trapped atoms up to a magnetic field gradient of about \unit{1.0}{\milli\tesla\per\centi\metre} shows a steep increase between \unit{0.3}{\milli\tesla\per\centi\metre} and  \unit{0.8}{\milli\tesla\per\centi\metre}.
The diameter of the atomic beam feeding the 3D MOT decreases for steeper magnetic field gradients and results in a better match to the small hole of the differential pumping stage.

\begin{figure}[H]
  \resizebox{.95\hsize}{!}{\includegraphics*{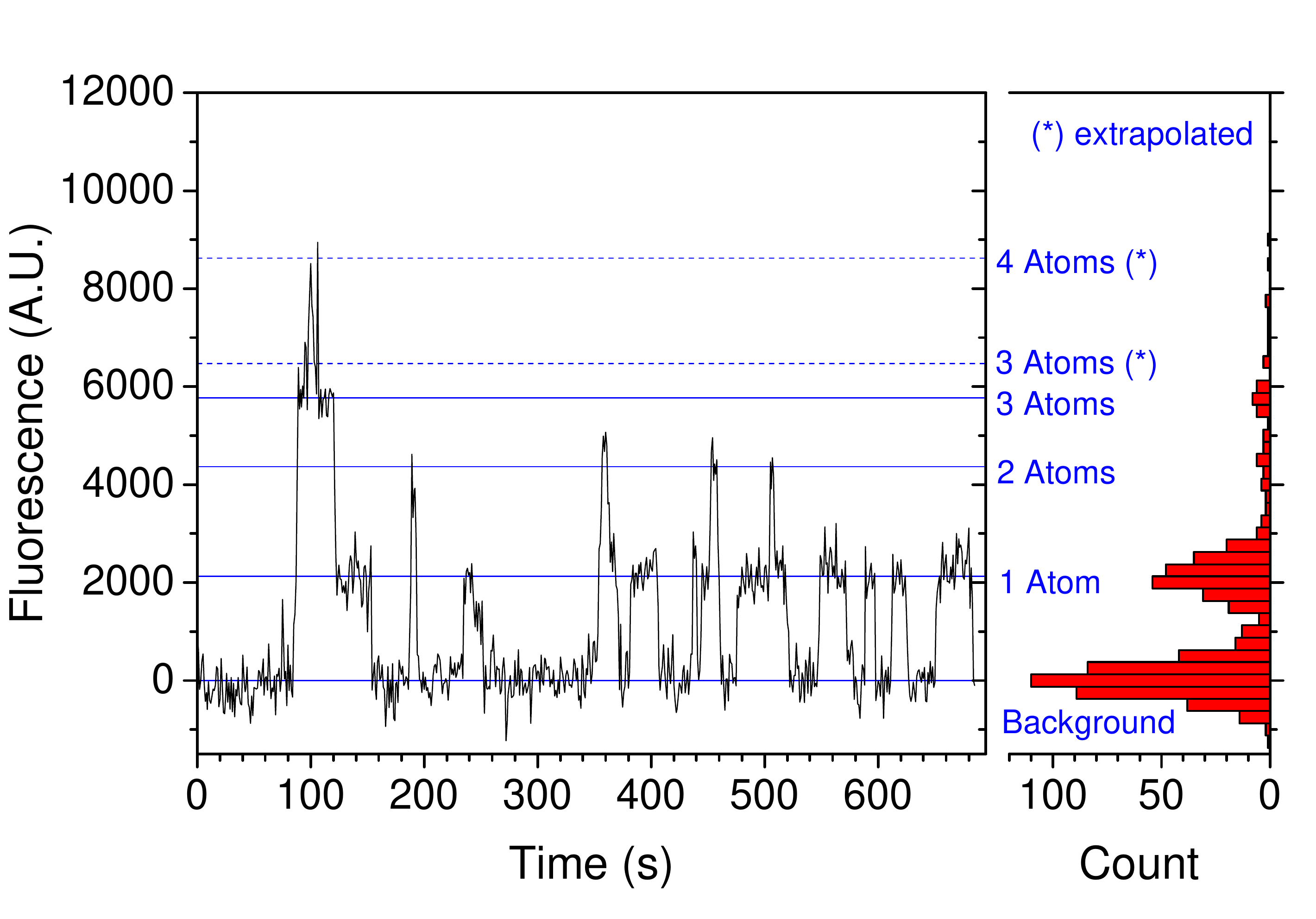}}
 
 {a) Up to four trapped $^{84}$Kr atoms.}
 
   \resizebox{.95\hsize}{!}{\includegraphics*{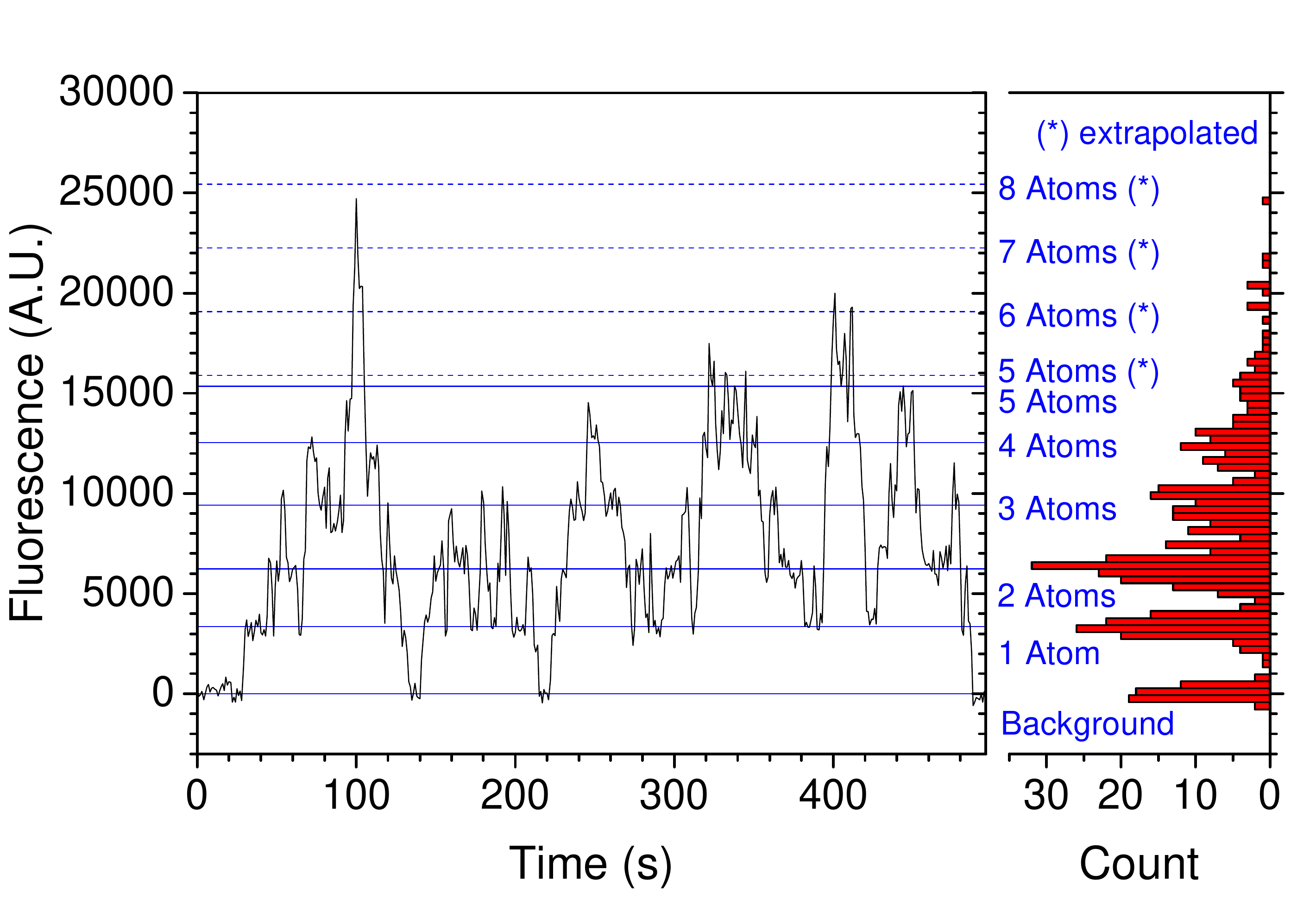}}

   {b) Up to eight trapped $^{84}$Kr atoms.}

   \resizebox{.95\hsize}{!}{\includegraphics*{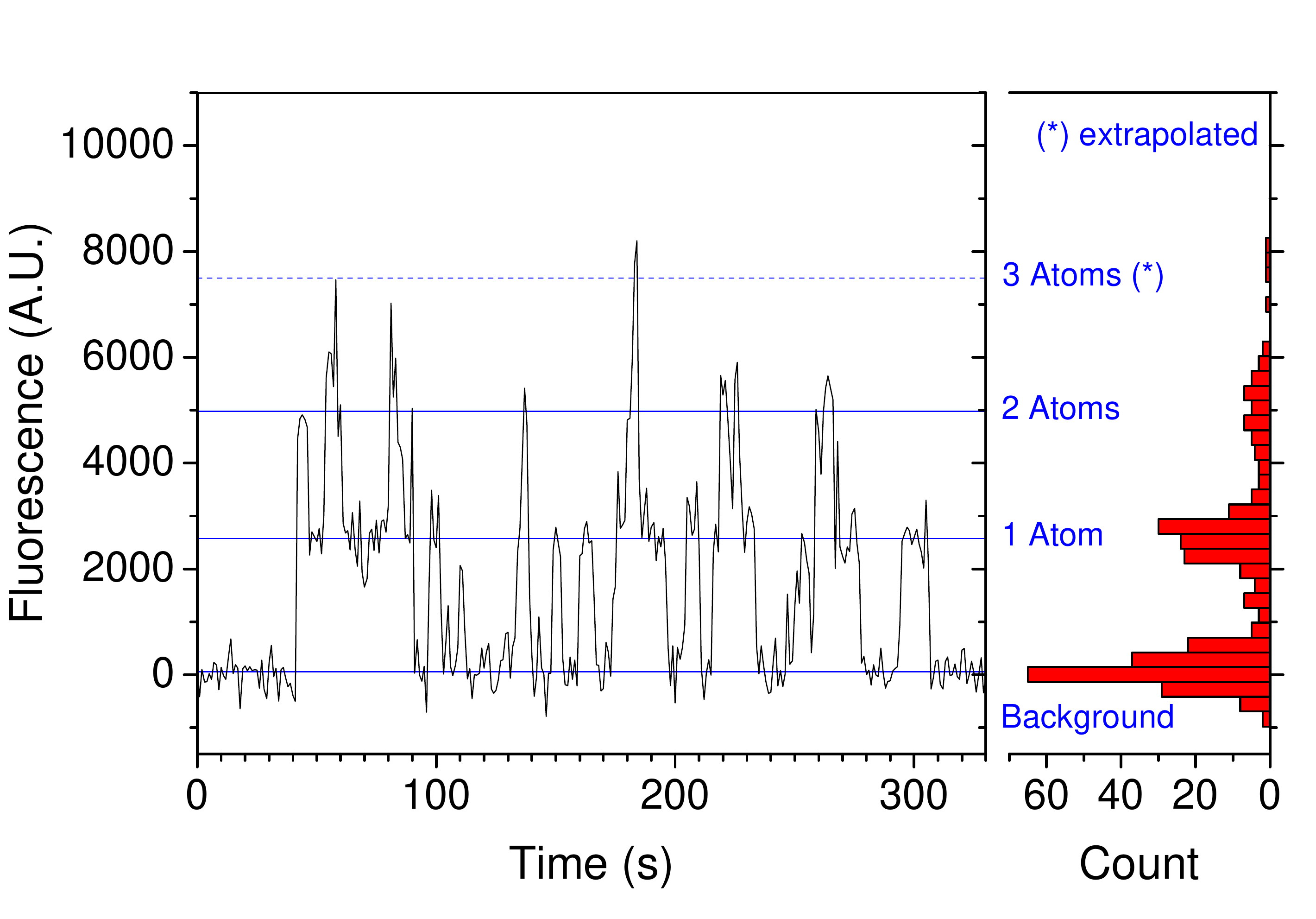}}
   
{c) Up to three trapped $^{83}$Kr atoms.}
 \caption{Fluorescence of single trapped $^{84}$Kr or $^{83}$Kr atoms as a function of time. 
 The experimental parameters were adjusted to a flux of metastable atoms consisting of only a few atoms per minute. 
 In all of the figures, the fluorescence levels corresponding to different atom numbers are plotted. 
 The dotted lines are multiples of one atom fluorescence level, while the full lines correspond to the measured mean of the fitted fluorescence distribution of individual atom numbers. 
 Integration time amounts to 1 s.\label{single}
}
\end{figure}

\textbf{Single atom detection}
The detection of single atoms captured within the 3D MOT is the feature which fully exploits the benefit of ATTA.
Single atom detection for neutral atoms was first demonstrated \cite{1996-haubrich-meschede-europhys-lett-v34-p663, 0295-5075-34-9-651} and results exhibit precise determination of atom numbers in mesoscopic samples \cite{hume2013accurate}.
To demonstrate the single atom detection capability of our setup, we reduced the flux of metastable atoms originating from the 2D MOT by lowering the pressure within the 2D MOT, 
the power of the \unit{819}{\nano\meter} radiation, and the microwave power given to the lamps until single atoms were observed in the 3D MOT.
We could demonstrate single atom detection for both isotopes, $^{84}$Kr and $^{83}$Kr.
In Figure~\ref{single} we observe very few atoms within the 3D MOT, ranging from one to four atoms or one to eight atoms. 
For large numbers of atoms, the measured mean fluorescence $f_{\bar{n}}$ becomes smaller than the intuitively expected value $n \times f_{\bar{1}}$, since the probability of finding the corresponding number of atoms $n$ over  the whole integration time of \unit{1}{\second} is smaller than 1.
To be more precise, the probability of collisions between trapped atoms increases as a function of $n$ as can be determined from Figure~\ref{single}.

\section{Conclusion}
We demonstrate the first neutral noble gas 2D--3D MOT apparatus with combined all-optical production of meta\-sta\-ble atoms and cooling.
This feature relies on the first successful implementation of a 2D MOT with VUV lamps that have a long lifetime.
Our results, based on optical production and single atom detection capability, pave the way to overcome the technical problems of present RF-driven noble gas trace analysis techniques.

Future ATTA setups using our new technology will allow higher sample throughput, and, equally as important, smaller sample sizes,
as it avoids cross-con\-tam\-i\-na\-tion and does not require large samples for sustaining a plasma.
Our new approach will further increase the applicability of noble gas trace analysis in physics, earth sciences and nuclear arms control.

\section*{Acknowledgments}
We gratefully acknowledge the technical support of the staff members of the mechanic workshop of the department of chemistry at the University of Hamburg in developing our three chamber ultra high vacuum system including eight VUV lamps.

We thank Niko Lehmkuhl for a novel laser setup, and we thank Simon Hebel for trailblazing discussions about future applications.
We thank Franziska Herrmann for designing the imaging optics.

We would like to thank the Deutsche Forschungsgemeinschaft (grant KA 821/1-1) and the German Foundation for Peace Research for funding.

 \bibliographystyle{plain}
\bibliography{paper.bib}

\begin{thebibliography}{10}

\bibitem{bardou1992magneto}
F.~Bardou, O.~Emile, J.-M. Courty, C.I. Westbrook, and A.~Aspect.
\newblock Magneto-optical trapping of metastable helium: collisions in the
  presence of resonant light.
\newblock {\em Europhysics Letters}, 20(8):681, 1992.

\bibitem{collon2004tracing}
P.~Collon, W.~Kutschera, and Z.-T. Lu.
\newblock {Tracing Noble Gas Radionuclides in the Environment}.
\newblock {\em Annu. Rev. Nucl. Part. Sci.}, 54:39--67, 2004.

\bibitem{daerr2011novel}
H.~Daerr, M.~Kohler, P.~Sahling, S.~Tippenhauer, A.~Arabi-Hashemi, C.~Becker,
  K.~Sengstock, and M.B. Kalinowski.
\newblock A novel vacuum ultra violet lamp for metastable rare gas experiments.
\newblock {\em Review of Scientific Instruments}, 82(7):073106, 2011.

\bibitem{du2003realization}
X.~Du.
\newblock {\em Realization of Radio-Krypton Dating With an Atom Trap}.
\newblock PhD thesis, Northwestern University, 2003.

\bibitem{1996-haubrich-meschede-europhys-lett-v34-p663}
D.~Haubrich, H.~Schadwinkel, F.~Strauch, B.~Ueberholz, R.~Wynands, and
  D.~Meschede.
\newblock {Observation of individual neutral atoms in magnetic and
  magneto-optical traps}.
\newblock {\em Europhysics Letters}, 34:663, 1996.

\bibitem{hume2013accurate}
D.B. Hume, I.~Stroescu, M.~Joos, W.~Muessel, H.~Strobel, and M.K. Oberthaler.
\newblock {Accurate Atom Counting in Mesoscopic Ensembles}.
\newblock {\em Physical Review Letters}, 111(25):253001, 2013.

\bibitem{kalinowski2006isotopic}
M.~B. Kalinowski and C.~Pistner.
\newblock Isotopic signature of atmospheric xenon released from light water
  reactors.
\newblock {\em Journal of environmental radioactivity}, 88(3):215--235, 2006.

\bibitem{kalinowski2004conclusions}
M.~B. Kalinowski, H.~Sartorius, S.~Uhl, and W.~Weiss.
\newblock {Conclusions on plutonium separation from atmospheric krypton-85
  measured at various distances from the Karlsruhe reprocessing plant}.
\newblock {\em Journal of environmental radioactivity}, 73(2):203--222, 2004.

\bibitem{katori1993lifetime}
H.~Katori and F.~Shimizu.
\newblock Lifetime measurement of the 1s5 metastable state of argon and krypton
  with a magneto-optical trap.
\newblock {\em Physical review letters}, 70:3545--3548, 1993.

\bibitem{lu2010atom}
Z.-T. Lu and P.~M{\"u}ller.
\newblock Atom trap trace analysis of rare noble gas isotopes.
\newblock {\em Advances In Atomic, Molecular, and Optical Physics},
  58:173--205, 2010.

\bibitem{lu2013tracer}
Z.-T. Lu, P.~Schlosser, W.M. Smethie~Jr, N.C. Sturchio, T.P. Fischer, B.M.
  Kennedy, R.~Purtschert, J.P. Severinghaus, D.K. Solomon, T.~Tanhua, et~al.
\newblock Tracer applications of noble gas radionuclides in the geosciences.
\newblock {\em Earth-Science Reviews}, 2013.

\bibitem{okabe1964intense}
H.~Okabe.
\newblock Intense resonance line sources for photochemical work in the vacuum
  ultraviolet region.
\newblock {\em JOSA}, 54(4):478--481, 1964.

\bibitem{0295-5075-34-9-651}
F.~Ruschewitz, D.~Bettermann, J.~L. Peng, and W.~Ertmer.
\newblock Statistical investigations on single trapped neutral atoms.
\newblock {\em Europhysics Letters}, 34(9):651, 1996.

\bibitem{shimizu1992double}
F.~Shimizu, K.~Shimizu, and H.~Takuma.
\newblock Double-slit interference with ultracold metastable neon atoms.
\newblock {\em Physical review. A}, 46(1):R17--R20, 1992.

\bibitem{sturchio2004one}
N.C. Sturchio, X.~Du, R.~Purtschert, B.E. Lehmann, M.~Sultan, L.J. Patterson,
  Z.-T. Lu, P.~M{\"u}l\-ler, T.~Big\-ler, K.~Bailey, et~al.
\newblock {One million year old groundwater in the Sahara revealed by
  krypton-81 and chlorine-36}.
\newblock {\em Geophysical Research Letters}, 31(5), 2004.

\bibitem{walhout1993magneto}
M.~Walhout, H.J.L. Megens, A.~Witte, and S.L. Rolston.
\newblock {Magneto-optical trapping of metastable xe\-non: Isotope-shift
  measurements}.
\newblock {\em Physical Review A}, 48(2):R879, 1993.

\bibitem{0953-4075-35-13-311}
L.~Young, D.~Yang, and R.W. Dunford.
\newblock Optical production of metastable krypton.
\newblock {\em Journal of Physics B: Atomic, Molecular and Optical Physics},
  35(13):2985, 2002.

\end{thebibliography}

\end{multicols}
\end{document}